\documentclass[runningheads]{llncs}

\usepackage{hyperref}
\usepackage{color}
\usepackage{xspace}

\usepackage[T1]{fontenc}
\usepackage{booktabs}
\usepackage{amsmath}
\usepackage{multirow}
\usepackage{enumitem}
\usepackage{amsmath}
\usepackage{amsfonts}
\usepackage{placeins}
\usepackage{float}
\usepackage{enumitem}

\usepackage{algorithm}
\usepackage{algorithmic}

\newcommand{\sstitle}[1]{\smallskip\noindent\textbf{#1.\/}}

\newcommand{\toolname}{\texttt{SemFGRec}\xspace}

\usepackage{epstopdf}
\usepackage{graphics}
\DeclareGraphicsExtensions{.eps,.gz,.eps,.pdf,.png,.jpg}
\graphicspath{{./}{./figures/}{../figures/}}
\usepackage{graphicx}
\begin{document}

\title{Improving Federated Graph Recommendation with Semantic Guidance}

\author{
Thi Minh Chau Nguyen
\and
Hien Trang Nguyen
\and
Duc Anh Nguyen
\and
Van Ho-Long
\and
Thanh Trung Huynh
\and
Zhao Ren
}

\institute{}

\maketitle

\begin{abstract}
\begin{sloppypar}
Graph-based recommendation models effectively capture high-order collaborative signals from user--item interaction graphs. Federated learning (FL) enables privacy-preserving training across distributed clients. However, directly aggregating graph representations under FL is challenging: locally learned structural embeddings are not globally aligned under non-IID data distributions, and naive parameter averaging fails to recover cross-client relational structure. Existing federated graph-based approaches primarily rely on structural aggregation, yet overlook the global semantic knowledge encoded in large language models (LLMs). In this work, we propose a semantic--structural federated graph recommendation framework that leverages LLM embeddings to guide cross-client alignment. Each client learns user representations from its local interaction graph and summarizes typical interaction patterns into compact semantic vectors using a frozen LLM encoder. These vectors are sent to the server, which identifies semantically related patterns across different clients and combines their structural representations accordingly. The updated representations are then returned to clients to refine subsequent local training. This design enables collaboration guided by shared semantic understanding without exposing raw interaction data, preserving both recommendation accuracy and privacy. Experiments on benchmark datasets demonstrate consistent improvements over existing federated graph-based baselines.
\end{sloppypar}

\end{abstract}

\setlength{\belowdisplayskip}{2pt}
\setlength{\belowdisplayshortskip}{2pt}
\setlength{\abovedisplayskip}{2pt}
\setlength{\abovedisplayshortskip}{2pt}

\section{Introduction}
\label{sec:introduction}

Personalized recommendation systems have become a core component of modern online services, including e-commerce, media streaming, and social platforms. Classical collaborative filtering techniques~\cite{koren2009matrix,he2017neural} learn latent user and item representations from large-scale interaction data stored in centralized servers. More recently, graph neural networks (GNNs) have demonstrated strong capability in modeling high-order user--item relationships by propagating information over interaction graphs, as exemplified by LightGCN~\cite{he2020lightgcn}. Despite their effectiveness, these centralized paradigms require collecting and storing raw user interaction logs, raising substantial privacy and regulatory concerns.

Federated learning (FL) offers a natural alternative by enabling collaborative model training without sharing raw data. In federated recommender systems, each client (e.g., user device or organization) retains its local interaction history and contributes model updates to a central server~\cite{ammad2019federated}. Subsequent work has improved federated recommendation from multiple perspectives, including handling heterogeneous client distributions~\cite{yuan2024hetefedrec}
FedRecon~\cite{singhal2021federated} and PFedRec~\cite{zhang2023dual} further adapt global models to client-specific preferences. However, most federated methods rely on parameter aggregation schemes that implicitly assume structural compatibility across clients.

Graph-based federated recommendation has emerged to better exploit relational structures under privacy constraints. Approaches such as FedPerGNN~\cite{wu2022federated}, GPFedRec~\cite{zhang2024gpfedrec}, UFGraphFR~\cite{wang2025ufgraphfr}, and GFed-PP~\cite{na2025graph} incorporate graph modeling or user-relation construction into federated optimization. These methods demonstrate that structural signals beyond independent user modeling can improve personalization. Nevertheless, they typically align models at the embedding or user-graph level, assuming that structurally similar patterns across clients are directly comparable. In realistic federated environments, client data are highly non-IID: user communities on different devices may share semantic preferences but exhibit divergent local graph structures. Direct structural averaging or graph aggregation may therefore fail to capture cross-client semantic consistency.

In parallel, large language models (LLMs) have shown remarkable ability to encode rich semantic information from textual descriptions and behavioral contexts~\cite{llmrec,wu2024llm4rec_survey,tang2025}. LLM-based recommendation frameworks demonstrate that semantic abstraction can complement ID-based collaborative signals, particularly when interaction data are sparse or heterogeneous. In federated settings, recent efforts such as GPT-FedRec~\cite{zeng2024federated} and FELLAS~\cite{yuan2025fellas} leverage LLMs to enhance representation learning or sequential modeling under privacy constraints. However, existing approaches primarily leverage LLMs to refine local representations or enrich item semantics, rather than employing them to guide structural alignment across distributed clients in federated environments.

These observations motivate a central question: \emph{can semantic abstraction be used to regulate structural collaboration in federated graph recommendation?} Instead of aggregating graph embeddings directly, we argue that cross-client alignment should first be established at the semantic level, and then modulate structural merging accordingly. Semantic similarity provides a more stable signal under heterogeneous distributions, as communities that share preference intent may not exhibit identical graph topology. By prioritizing semantic alignment and using structural similarity as a secondary modulation, federated collaboration can become both more robust and more interpretable.

In this work, we propose \toolname{} (\emph{Semantic Federated Graph Recommendation}), a semantic-guided federated graph learning framework designed for non-IID environments. \toolname operates by learning collaborative representations from local interaction graphs using a lightweight graph encoder~\cite{he2020lightgcn}, abstracting preference patterns into higher-level descriptors, and enabling semantically consistent knowledge sharing across clients without exposing raw interaction data. Instead of blindly aggregating structural parameters, \toolname introduces a semantic-aware collaboration mechanism that selectively regularizes structurally compatible representations across clients. This design preserves personalization while enabling principled cross-client knowledge transfer.

In summary, our contributions are threefold:
\begin{itemize}[noitemsep]
    \item We introduce \toolname, a semantic-guided framework for federated graph recommendation that uses semantic signals to regulate how structural patterns are shared across clients. 
    \item We design a semantic-aware collaboration mechanism improving robustness under non-IID client distributions without transmitting raw interaction data.
    \item We conduct comprehensive experiments against strong centralized, federated, and graph-based baselines, demonstrating consistent gains and improved stability across heterogeneous environments.
\end{itemize}
\section{Related Works}
\label{sec:related}

\sstitle{Federated Recommender Systems}
Federated recommendation has emerged to reconcile personalization with strict data protection requirements~\cite{yin2025device,long2024diffusion}. 
Subsequent research improved optimization under heterogeneous client distributions~\cite{yuan2024hetefedrec}, strengthened privacy guarantees through differential privacy and secure aggregation
~\cite{long2024physical,qu2024towards,yuan2024hide}
PFedRec~\cite{zhang2023dual} introduces dual personalization mechanisms for both users and items. Meta-learning and lightweight parameter adaptation have also been explored to improve scalability. Extensions include decentralized aggregation without a single central server (DeFedGCN)~\cite{chen2025defedgcn}, and privacy-preserving knowledge-graph-aware recommendation frameworks such as FedKGRec~\cite{ma2024fedkgrec}. Cross-domain federated graph transfer is studied in FedGCDR~\cite{yang2024federated}. Explainable and knowledge-graph-enhanced federated frameworks further improve transparency and interpretability~\cite{kumar2025depth,soyarar2025explaining}. Despite these advances, existing methods primarily emphasize collaborative optimization, privacy mechanisms, or personalization, without explicitly incorporating semantic abstraction for cross-client structural alignment.

\sstitle{Federated Recommender Systems with LLM}
Large language models (LLMs) have recently been integrated into recommender systems to enhance representation learning and generative capability~\cite{llmrec,wu2024llm4rec_survey}. Some studies employ LLMs as domain-agnostic recommenders~\cite{tang2025}, while others combine retrieval-augmented generation with federated learning (e.g., GPT-FedRec~\cite{zeng2024federated}). Federated LLM-based recommendation frameworks include FELLRec~\cite{zhao2025federated} and FELLAS~\cite{yuan2025fellas}. Broader perspectives on federated LLM training are discussed in recent surveys and benchmarks~\cite{yao2024,ye2024}. In centralized settings, multi-behavior graph modeling such as MBH-GNN~\cite{tan2025nah} demonstrates that behavior-aware semantic embeddings can improve recommendation accuracy. However, most LLM-based federated approaches focus on representation enhancement or re-ranking, rather than using semantic abstractions to regulate structural collaboration across distributed clients in federated environments.

\sstitle{Federated Recommender Systems with Graph}
Graph neural networks (GNNs) have become a dominant paradigm for recommendation due to their ability to capture high-order user--item interactions. LightGCN~\cite{he2020lightgcn} simplifies graph convolution to linear neighborhood aggregation and serves as an effective backbone for collaborative filtering. In federated environments, FeSoG~\cite{liu2022federated} extends graph modeling to social recommendation under privacy constraints. FedPerGNN~\cite{wu2022federated} enables decentralized graph learning with privacy-preserving graph expansion. GPFedRec~\cite{zhang2024gpfedrec} constructs user-relation graphs from personalized item embeddings to guide aggregation. UFGraphFR~\cite{wang2025ufgraphfr} incorporates semantic similarity to build federated user graphs. GFed-PP~\cite{na2025graph} leverages both private and public user data for privacy-aware aggregation. Subgraph-level privacy-preserving learning is explored in PFGRS~\cite{qi2025pfgrs}. Vertical federated graph training is studied in VerFedGNN~\cite{mai2023vertical}. Additional enhancements integrate structured clustering and graph attention within federated learning frameworks~\cite{xu2025enhancing}. Although these approaches combine graph modeling with federated optimization, they operate primarily at the embedding or user-graph level and do not perform semantic-guided community-level structural merging.

Overall, prior research addresses collaborative optimization, graph modeling, personalization, knowledge graphs, and LLM-based enhancement largely in isolation. In contrast, our framework integrates semantic abstraction and structural graph modeling at the community level, enabling semantic-first alignment to improve robustness under heterogeneous federated environments.
\section{Problem Formulation}
\label{sec:problem}

We consider a federated graph-based recommendation setting with $K$ clients and a central server. 
Each client $k \in \{1,\dots,K\}$ holds a local user--item interaction graph 
\[
\mathcal{G}_k = (\mathcal{U}_k, \mathcal{I}_k, \mathcal{E}_k),
\]
where $\mathcal{U}_k$ denotes the set of local users, $\mathcal{I}_k \subseteq \mathcal{I}$ denotes the items observed by client $k$, and $\mathcal{E}_k \subseteq \mathcal{U}_k \times \mathcal{I}_k$ represents observed interactions (e.g., clicks, purchases, or ratings).

The global item set is denoted by $\mathcal{I}$ with $|\mathcal{I}| = M$. 
Users are disjoint across clients, i.e., $\mathcal{U}_k \cap \mathcal{U}_{k'} = \emptyset$ for $k \neq k'$, while items may be shared. 
Raw interaction data $\mathcal{E}_k$ and user identities are kept on-device and are not transmitted to the server, reducing direct exposure of user behavior.

\subsection{Federated Objective}

The goal is to learn client-specific user embeddings $\{h_u \in \mathbb{R}^d \mid u \in \mathcal{U}_k\}$ for each client $k$ and a shared global item embedding matrix $E_I \in \mathbb{R}^{M \times d}$ such that recommendation quality is maximized while preserving data locality.

For each client $k$, we adopt the Bayesian Personalized Ranking (BPR) objective as the local optimization target:

\begin{equation}
\mathcal{L}_{\text{rank}}^k
=
- \mathbb{E}_{(u,i^+,i^-)\sim \mathcal{D}_k}
\log \sigma\!\left(
\langle h_u, h_{i^+} \rangle
-
\langle h_u, h_{i^-} \rangle
\right),
\end{equation}
where $\mathcal{D}_k$ denotes the local training distribution constructed from $\mathcal{E}_k$.
Here, $(u,\allowbreak i^+,\allowbreak i^-)$ represents a user, a positive item, and a negative item, respectively.

In a standard federated learning framework, the global objective can be written as:

\begin{equation}
\min_{\{h_u\}, E_I}
\sum_{k=1}^K
\mathcal{L}_{\text{rank}}^k,
\end{equation}
subject to the constraint that raw interaction data remain local.

\subsection{Challenges Under Non-IID Graph Distributions}

In practice, client data distributions are non-IID. 
Different clients may exhibit distinct preference domains, interaction densities, and item popularity patterns. 
Under such heterogeneity, naive federated averaging of model parameters does not guarantee alignment of learned representations.

In particular, user embeddings are client-specific and cannot be aggregated directly. 
Even for shared item embeddings, simple averaging may mix incompatible preference signals across clients. 
Moreover, graph structures learned locally may encode structurally similar patterns that correspond to semantically different user intents.

\subsection{Semantic--Structural Alignment Objective}

To address these challenges, we introduce community-level representations on each client. 
Let $\{\mathcal{U}_{k,c}\}_{c=1}^{C_k}$ denote preference communities identified on client $k$. 
For each community, we compute:

\begin{itemize}
    \item a structural prototype $P_{k,c} \in \mathbb{R}^d$, summarizing collaborative preference geometry, and
    \item a semantic embedding $S_{k,c} \in \mathbb{R}^d$, summarizing thematic characteristics derived from item content.
\end{itemize}

Our objective extends the standard federated ranking loss by introducing a cross-client alignment term:

\begin{equation}
\mathcal{L}_k
=
\mathcal{L}_{\text{rank}}^k
+
\lambda_{\text{align}}
\sum_{c=1}^{C_k}
\left\|
P_{k,c}
-
\tilde{P}_g^{(t)}
\right\|_2^2,
\end{equation}
where $\tilde{P}_g^{(t)}$ denotes the merged structural prototype of semantically aligned communities across clients.

The key idea is to perform structural collaboration only when semantic consistency exists. 
By aligning community prototypes rather than individual user embeddings, we enable knowledge transfer across clients while maintaining personalization and privacy.

In summary, the problem is to design a federated optimization procedure that:

\begin{itemize}
    \item learns collaborative representations from local interaction graphs,
    \item identifies semantically coherent preference communities,
    \item merges structurally compatible communities across clients, and
    \item improves recommendation accuracy without sharing raw interaction data.
\end{itemize}
\section{Methodology}
\label{sec:methodology}

\begin{figure*}[t]
    \centering
    \includegraphics[width=\textwidth]{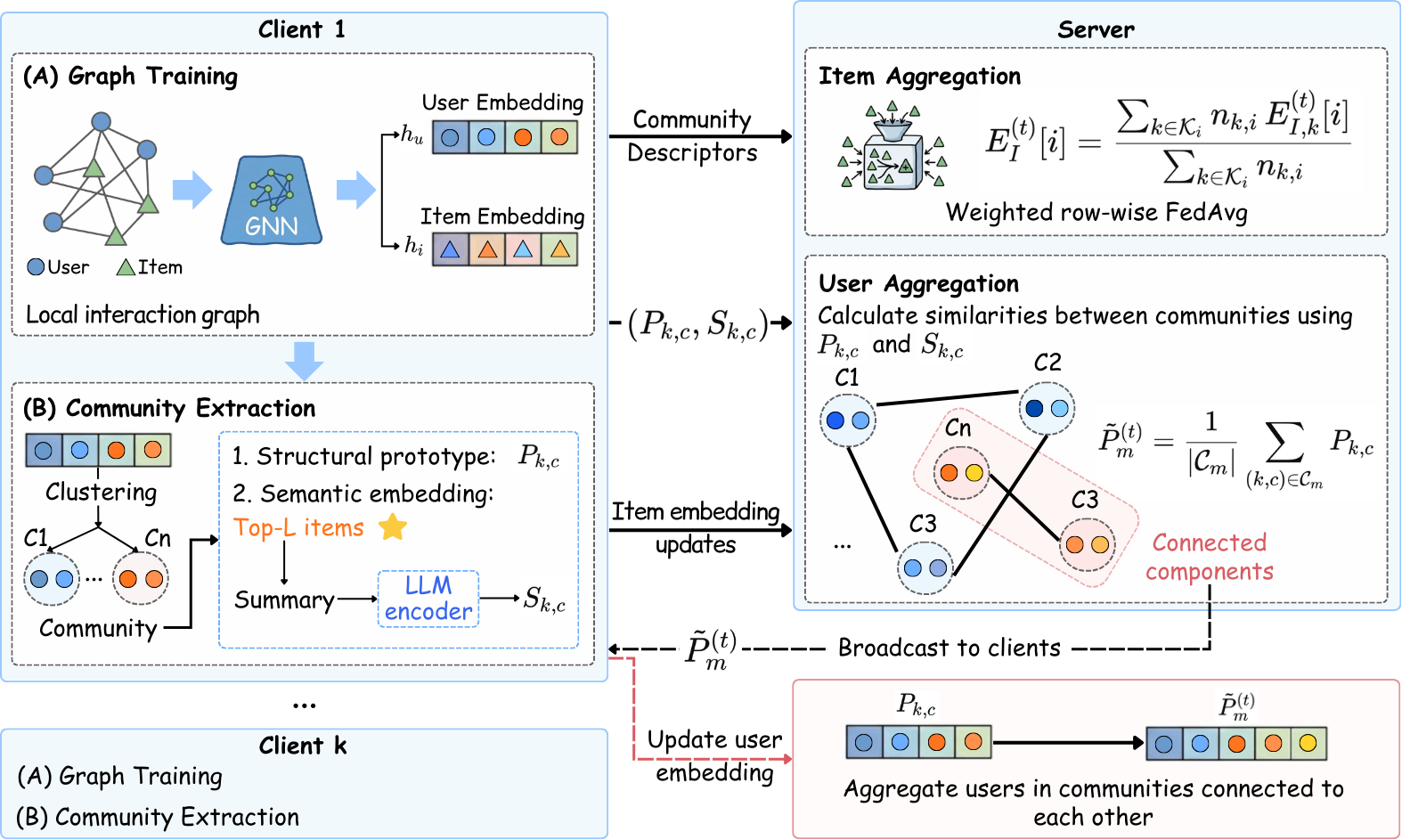}
    \caption{Overall framework of \toolname. Each client performs local graph-based representation learning and adaptive community abstraction, producing structural prototypes $P_{k,c}$ and semantic descriptors $S_{k,c}$. The server aggregates item embeddings and constructs a semantic-guided community graph to merge aligned communities, generating global prototypes that are broadcast back for alignment-aware local refinement.}
    \label{fig:framework}
\end{figure*}

Figure~\ref{fig:framework} illustrates the overall pipeline of \toolname. At each communication round, clients first learn user and item representations from their local interaction graphs using a lightweight GNN encoder. Based on the learned user embeddings, each client extracts preference communities and derives two complementary descriptors: a structural prototype $P_{k,c}$ summarizing collaborative patterns and a semantic embedding $S_{k,c}$ obtained by encoding the summarized metadata of representative items. The server then performs row-wise item aggregation and constructs a semantic-first community graph across clients, where structural similarity modulates semantic alignment. Communities that satisfy the merging criterion are grouped and averaged to form global prototypes, which are broadcast back to clients for alignment-aware optimization. This iterative procedure enables cross-client semantic consistency while preserving local personalization under federated constraints.

\subsection{Local Collaborative Representation Learning}

On each client, we learn user and item representations from the local user--item interaction graph using a LightGCN-style encoder \cite{he2020lightgcn}. 
Given the normalized adjacency matrix $\tilde{A}_k$, embeddings are propagated for $L_g$ layers:

\begin{equation}
H_k^{(l+1)} = \tilde{A}_k H_k^{(l)}.
\end{equation}

The final embedding of user $u$ and item $i$ is obtained by aggregating representations from all layers:

\begin{equation}
h_u = \sum_{l=0}^{L_g} \alpha_l h_u^{(l)}, 
\qquad
h_i = \sum_{l=0}^{L_g} \alpha_l h_i^{(l)},
\end{equation}
where $\alpha_l = \frac{1}{L_g+1}$ in our implementation. 
This encoder captures high-order collaborative signals through neighborhood propagation while keeping the architecture simple. 
A lightweight design is preferable in federated settings because it reduces training variance across clients and makes the contribution of the semantic--structural merging mechanism easier to isolate.

Each client first minimizes the local ranking objective 
$\mathcal{L}_{\text{rank}}^k$ defined in Section~\ref{sec:problem}:

\begin{equation}
\mathcal{L}_{\text{rank}}^k
=
- \mathbb{E}_{(u,i^+,i^-)\sim \mathcal{D}_k}
\log \sigma\!\left(
\langle h_u, h_{i^+}\rangle
-
\langle h_u, h_{i^-}\rangle
\right),
\end{equation}
which encourages observed interactions to receive higher scores than sampled negatives.

\subsection{Adaptive Community Abstraction}

After learning user embeddings $\{h_u\}$, each client partitions users into preference communities using HDBSCAN \cite{campello2013density}, a density-based clustering algorithm:

\begin{equation}
\{\mathcal{U}_{k,c}\}_{c=1}^{C_k}
=
\texttt{HDBSCAN}(\{h_u\}).
\end{equation}

HDBSCAN determines the number of clusters automatically based on embedding density. 
This is important in federated environments, where different clients may exhibit different levels of interaction sparsity and preference diversity. 
Fixing the number of communities would impose an artificial structural assumption that may not reflect local data characteristics.

For each community $c$, we compute a structural prototype:

\begin{equation}
P_{k,c}
=
\frac{1}{|\mathcal{U}_{k,c}|}
\sum_{u \in \mathcal{U}_{k,c}} h_u.
\end{equation}

This prototype summarizes the collaborative preference pattern of the community in embedding space.

To obtain a semantic abstraction, we first select the top-$L$ items with the highest interaction frequency among users in $\mathcal{U}_{k,c}$. 
We then aggregate their textual metadata (e.g., titles and descriptions) to construct a community-level summary $x_{k,c}$, which is encoded using a frozen large language model:

\begin{equation}
S_{k,c}
=
\text{LLMEnc}(x_{k,c}).
\end{equation}

The structural prototype $P_{k,c}$ captures interaction-driven similarity, whereas the semantic embedding $S_{k,c}$ captures thematic similarity derived from item content. 
These two representations provide complementary views of the same community.

\subsection{Semantic-Guided Structural Merging}

The server first aggregates item embeddings via row-wise weighted averaging:

\begin{equation}
E_I^{(t)}[i]
=
\frac{\sum_k n_{k,i} E_{I,k}[i]}
{\sum_k n_{k,i}}.
\end{equation}

To align communities across clients, we compute semantic similarity using cosine similarity:

\begin{equation}
s_{a,b}
=
\frac{S_a^\top S_b}{\|S_a\|\|S_b\|}.
\end{equation}

Cosine similarity is appropriate because both LLM embeddings and collaborative embeddings encode information primarily in their direction rather than magnitude.

Structural similarity is computed similarly:

\begin{equation}
r_{a,b}
=
\frac{P_a^\top P_b}{\|P_a\|\|P_b\|}.
\end{equation}

We combine the two similarities as:

\begin{equation}
q_{a,b}
=
s_{a,b}\left(1 + \lambda r_{a,b}\right).
\end{equation}

In this formulation, semantic similarity acts as the primary alignment signal. 
Structural similarity only adjusts the strength of merging. 
This design prevents merging communities that exhibit similar interaction statistics but correspond to different semantic meanings. 
In other words, semantic coherence is required before structural compatibility is considered.

Communities are merged when $q_{a,b} > \delta$, and merged structural prototypes are computed as:

\begin{equation}
\tilde{P}_g^{(t)}
=
\frac{1}{|\mathcal{C}_g|}
\sum_{a \in \mathcal{C}_g} P_a.
\end{equation}

\subsection{Alignment-Aware End-to-End Training}

After receiving merged prototypes, each client optimizes the following alignment-aware objective:

\begin{equation}
\mathcal{L}_k
=
\mathcal{L}_{\text{rank}}^k
+
\lambda_{\text{align}}
\sum_{c=1}^{C_k}
\left\|
P_{k,c}
-
\tilde{P}_g^{(t)}
\right\|_2^2.
\end{equation}

The first term preserves personalized ranking performance.
The second term encourages structurally similar communities to remain aligned across clients when semantic consistency exists.
We use a soft alignment term rather than directly replacing local prototypes. 
Hard replacement may override locally meaningful preference patterns, especially under non-IID conditions. 
The alignment loss encourages consistency with semantically aligned communities while preserving local personalization.

Across communication rounds, we jointly optimize collaborative representation learning and semantic-guided merging.
This end-to-end process integrates structural interaction modeling with global semantic alignment, enabling effective federated recommendation while keeping raw interaction data local, and reducing direct exposure of user behavior.

\subsection{Training Procedure}

The overall procedure consists of two coordinated components executed at each communication round: 
(i) client-side structural learning with adaptive community abstraction, and 
(ii) server-side semantic-guided structural merging. 
Unlike fixed-cluster approaches, the number of communities is automatically determined per client, enabling adaptation to heterogeneous non-IID preference distributions.

\paragraph{Client-Side Procedure.}

Each client $k$ performs collaborative representation learning using a lightweight bipartite graph encoder, followed by adaptive community detection in embedding space and semantic abstraction. The detailed procedure is described in Algorithm~\ref{alg:client}.

\begin{algorithm}[!ht]
\caption{Client-Side Procedure at Round $t$}
\label{alg:client}
\begin{algorithmic}[1]
\REQUIRE Global item embeddings $E_I^{(t-1)}$, local interaction graph $\mathcal{G}_k$, local epochs $T_{\text{local}}$, alignment epochs $T_{\text{align}}$, top-$L$ representative items, alignment weight $\lambda_{\text{align}}$
\STATE Initialize local item embeddings $E_{I,k} \leftarrow E_I^{(t-1)}$

\STATE \textbf{(A) Collaborative structural learning (LightGCN-style)}
\FOR{$e = 1$ to $T_{\text{local}}$}
    \STATE Obtain user/item embeddings $\{h_u\},\{h_i\}$ via a LightGCN-style encoder on $\mathcal{G}_k$
    \STATE Update local parameters by minimizing the BPR ranking loss $\mathcal{L}_{\text{rank}}$
\ENDFOR

\STATE \textbf{(B) Adaptive community detection and semantic descriptors}
\STATE Cluster user embeddings $\{h_u\}$ via density-based clustering to obtain communities $\{\mathcal{U}_{k,c}\}_{c=1}^{C_k}$, where $C_k$ is automatically determined
\FOR{each community $c \in \{1,\dots,C_k\}$}
    \STATE Compute structural prototype $P_{k,c} \leftarrow \frac{1}{|\mathcal{U}_{k,c}|}\sum_{u \in \mathcal{U}_{k,c}} h_u$
    \STATE Select representative items $I^{\text{top}}_{k,c} \leftarrow \texttt{TopItems}(\mathcal{U}_{k,c}, L)$ and summarize $x_{k,c} \leftarrow \texttt{SummarizeItems}(I^{\text{top}}_{k,c})$
    \STATE Compute semantic embedding $S_{k,c} \leftarrow \texttt{LLMEnc}(x_{k,c})$
\ENDFOR
\STATE Upload sparse item updates and community descriptors $\{(S_{k,c}, P_{k,c})\}_{c=1}^{C_k}$

\STATE \textbf{(C) Post-merge structural alignment}
\FOR{$e = 1$ to $T_{\text{align}}$}
    \STATE Recompute prototypes $\{P_{k,c}\}$ and retrieve the merged prototype $\tilde{P}_g^{(t)}$ for each local community
    \STATE Minimize $\mathcal{L}_k = \mathcal{L}_{\text{rank}} + \lambda_{\text{align}}\sum_{c=1}^{C_k}\left\|P_{k,c}-\tilde{P}_g^{(t)}\right\|_2^2$
\ENDFOR
\end{algorithmic}
\end{algorithm}

\paragraph{Server-Side Procedure.}

The server performs row-wise item aggregation and constructs a semantic-first community graph across all clients. Structural similarity modulates semantic alignment to prevent incompatible merges. Since each client uploads only sparse item updates and a compact set of community descriptors rather than raw interaction logs, the communication cost is bounded by the number of updated items and discovered communities on that client. On the server side, the additional processing scales with the number of received community descriptors, making the procedure practical for federated deployment. The detailed procedure is described in Algorithm~\ref{alg:server}.

\begin{algorithm}[!ht]
\caption{Server-Side Procedure at Round $t$}
\label{alg:server}
\begin{algorithmic}[1]
\REQUIRE Client uploads $\{\Delta E_{I,k}\}$ and $\{(S_{k,c},P_{k,c})\}$, modulation weight $\lambda$, threshold $\delta$, top-$K_{\text{nn}}$ neighbors

\STATE \textbf{(A) Row-wise item aggregation}
\FOR{each updated item $i$}
    \STATE Update $E_I^{(t)}[i] \leftarrow 
    \frac{\sum_k n_{k,i} E_{I,k}[i]}{\sum_k n_{k,i}}$
\ENDFOR

\STATE \textbf{(B) Semantic-guided community graph construction}
\STATE Let $\mathcal{V} = \{(k,c)\}$ denote all received communities
\FOR{candidate pairs $(a,b)$ from top-$K_{\text{nn}}$ semantic neighbors}
    \STATE Compute semantic similarity 
    $s_{a,b} \leftarrow \frac{S_a^\top S_b}{\|S_a\|\|S_b\|}$
    \STATE Compute structural similarity 
    $r_{a,b} \leftarrow \frac{P_a^\top P_b}{\|P_a\|\|P_b\|}$
    \STATE Compute combined similarity 
    $q_{a,b} \leftarrow s_{a,b}(1+\lambda r_{a,b})$
    \STATE Add edge $(a,b)$ if $q_{a,b} > \delta$
\ENDFOR

\STATE Identify connected components $\{\mathcal{C}_g\}_{g=1}^{G_t}$

\STATE \textbf{(C) Prototype merging}
\FOR{each component $\mathcal{C}_g$}
    \STATE Compute merged prototype 
    $\tilde{P}_g^{(t)} \leftarrow 
    \frac{1}{|\mathcal{C}_g|}\sum_{a \in \mathcal{C}_g} P_a$
\ENDFOR

\STATE Broadcast updated item embeddings $\{E_I^{(t)}\}$ and merged prototypes $\{\tilde{P}_g^{(t)}\}$

\end{algorithmic}
\end{algorithm}
\section{Experiments}
\label{sec:experiments}

We conduct extensive experiments to evaluate the effectiveness and robustness of the proposed semantic--structural federated graph recommendation framework. Specifically, we aim to answer the following research questions:

\begin{itemize}
    \item \textbf{RQ1:} Does the proposed method outperform strong centralized and federated recommendation baselines?
    \item \textbf{RQ2:} Is the proposed semantic--structural merging robust under varying degrees of non-IID data heterogeneity?
    \item \textbf{RQ3:} What is the contribution of each component in the framework?
    \item \textbf{RQ4:} Is the proposed method stable with respect to key hyperparameters?
\end{itemize}

\subsection{Experimental Setup}

\sstitle{Datasets}
We evaluate on three benchmark datasets: MovieLens-100K and MovieLens-1M~\cite{harper2015movielens}, as well as Amazon Video~\cite{ni2019justifying}. MovieLens-100K contains 100,000 ratings from 943 users on 1,682 movies, while MovieLens-1M includes one million ratings from 6,040 users on 3,952 movies. Amazon Video provides user interactions together with textual metadata such as item titles and descriptions, and is considerably sparser than MovieLens. For all datasets, ratings are converted to implicit feedback, and users or items with fewer than five interactions are removed.

\sstitle{Evaluation Protocol}
We adopt a leave-one-out evaluation protocol in which the most recent interaction of each user is held out for testing. For each test user, we rank the ground-truth item among 100 sampled negative items. Performance is measured using Hit Ratio at rank 10 (HR@10) and Normalized Discounted Cumulative Gain at rank 10 (NDCG@10) \cite{tan2025nah}. HR@10 measures whether the ground-truth item appears in the top-10 list, while NDCG@10 further considers ranking positions. All results are reported in percentage (\%). To simulate federated learning, users are partitioned into $K$ clients. We consider IID partition and non-IID partition. For non-IID simulation, we adopt a Dirichlet distribution \cite{yurochkin2019bayesian} with concentration parameter $\alpha \in \{0.1, 0.3, 1.0\}$. Smaller $\alpha$ corresponds to more severe data heterogeneity, while $\alpha=1.0$ represents moderate skew.

\sstitle{Baselines}
We compare the proposed method against centralized, federated non-graph, and federated graph recommendation models.

\textbf{Centralized Models.}
We first consider classical collaborative filtering baselines trained in a centralized manner. 
\textbf{MF} \cite{koren2009matrix} learns low-dimensional latent representations for users and items and models their interactions via inner products. 
\textbf{NCF} \cite{he2017neural} replaces the linear interaction in MF with a multi-layer perceptron to capture non-linear user--item relationships. 
\textbf{LightGCN} \cite{he2020lightgcn} propagates embeddings on the user--item bipartite graph without feature transformation, effectively modeling high-order collaborative signals.

\textbf{Federated Non-Graph Models.}
To evaluate the impact of federated learning without graph modeling, we include several collaborative baselines under distributed training. 
\textbf{FedMF} \cite{chai2020secure} extends matrix factorization to the federated setting, where user representations are updated locally and item parameters are aggregated globally. 
\textbf{FedNCF} \cite{perifanis2022federated} adapts neural collaborative filtering to decentralized environments by sharing item-related parameters while keeping user embeddings local. 
\textbf{PFedRec} \cite{zhang2023dual} introduces personalization by separating shared and client-specific components. 
\textbf{FedRecon} \cite{singhal2021federated} reconstructs local representations during each communication round to enhance personalization under heterogeneous data.

\textbf{Federated Graph Models.}
We further compare with state-of-the-art federated graph recommendation approaches. 
\textbf{FedLightGCN} extends LightGCN \cite{he2020lightgcn} to federated learning via parameter aggregation across clients. 
\textbf{GPFedRec} \cite{zhang2024gpfedrec} leverages graph-guided aggregation strategies to improve cross-client structural knowledge transfer. 
\textbf{UFGraphFR} \cite{wang2025ufgraphfr} integrates textual user features to construct similarity-aware user graphs in federated environments. 
\textbf{GFed-PP} \cite{na2025graph} focuses on personalized federated graph learning to mitigate representation drift under non-IID settings.

\sstitle{Implementation Details}
Embedding dimension is set to $d=64$, and the number of graph propagation layers is $L_g=2$. We use the Adam optimizer with learning rate $10^{-3}$. Each communication round consists of five local training epochs. Adaptive community abstraction is performed using HDBSCAN with minimum cluster size of 10. The number of representative items per community is set to $L=10$. 
The semantic encoder is a frozen pretrained language model, and semantic embeddings are $\ell_2$-normalized before similarity computation. We keep the semantic encoder fixed in order to avoid additional synchronization and communication overhead during federated training.
Hyperparameters $\lambda$, $\lambda_{\text{align}}$, and $\delta$ are tuned using validation sets.

\subsection{RQ1: Overall Performance}

Table~\ref{tab:main_results} presents the overall performance under the non-IID setting ($\alpha=0.3$).
Among all baselines, UFGraphFR achieves the strongest results: 75.7\% (HR@10) and 47.0\% (NDCG@10) on MovieLens-100K; 75.2\% and 46.0\% on MovieLens-1M; and 81.8\% and 72.6\% on Amazon-Video.
In comparison, \toolname consistently delivers the best performance, 
achieving 77.7\% and 49.0\%, 
77.3\% and 48.0\%, 
and 83.5\% and 74.3\% on the respective datasets. 
These correspond to absolute improvements of approximately two percentage points in both metrics over the strongest baseline, 
demonstrating superior robustness under heterogeneous client distributions.

\begin{table}[!t]
\centering
\caption{Overall performance comparison under the non-IID setting ($\alpha=0.3$). 
Best results are in \textbf{bold} and the best baseline results are underlined.}
\label{tab:main_results}
\begin{tabular}{l l cc cc cc}
\toprule
\multirow{2}{*}{Group} & \multirow{2}{*}{Method}
& \multicolumn{2}{c}{MovieLens-100K}
& \multicolumn{2}{c}{MovieLens-1M}
& \multicolumn{2}{c}{Amazon-Video} \\
\cmidrule(lr){3-4}\cmidrule(lr){5-6}\cmidrule(lr){7-8}
& & HR@10 & NDCG@10 & HR@10 & NDCG@10 & HR@10 & NDCG@10 \\
\midrule

\multirow{3}{*}{CenRec}
& MF            & 64.6 & 38.8 & 68.9 & 41.7 & 47.2 & 30.4 \\
& NCF           & 64.9 & 38.3 & 64.7 & 38.3 & 60.6 & 39.2 \\
& LightGCN      & 64.3 & 37.4 & 60.9 & 33.9 & 60.4 & 39.3 \\
\midrule

\multirow{9}{*}{FedRec}
& FedMF         & 65.1 & 39.3 & 68.0 & 41.0 & 59.7 & 38.6 \\
& FedNCF        & 60.9 & 34.2 & 60.7 & 34.3 & 57.9 & 36.9 \\
& PFedRec       & 71.9 & 43.6 & 73.3 & 44.3 & 59.6 & 37.8 \\
& FedRecon      & 64.8 & 38.3 & 63.5 & 38.8 & 58.8 & 38.1 \\
& MetaMF        & 66.7 & 41.1 & 46.0 & 25.6 & 58.0 & 37.6 \\
& FedLightGCN   & 23.8 & 12.7 & 36.7 & 14.9 & 56.1 & 34.9 \\
& GPFedRec      & 72.5 & 43.4 & 72.7 & 43.9 & 58.6 & 38.4 \\
& GFedPP        & 73.3 & 43.9 & 73.5 & 44.4 & \underline{82.2} & \underline{73.0} \\
& UFGraphFR    & \underline{75.7} & \underline{47.0} 
                 & \underline{75.2} & \underline{46.0} 
                 & 81.8 & 72.6 \\
\midrule

\multirow{1}{*}{Ours}
& \toolname     & \textbf{77.7} & \textbf{49.0} 
                & \textbf{77.3} & \textbf{48.0} 
                & \textbf{83.5} & \textbf{74.3} \\
\bottomrule
\end{tabular}
\end{table}

\subsection{RQ2: Robustness Under Non-IID Settings}

\begin{figure*}[!t]
    \centering
    \includegraphics[width=\textwidth]{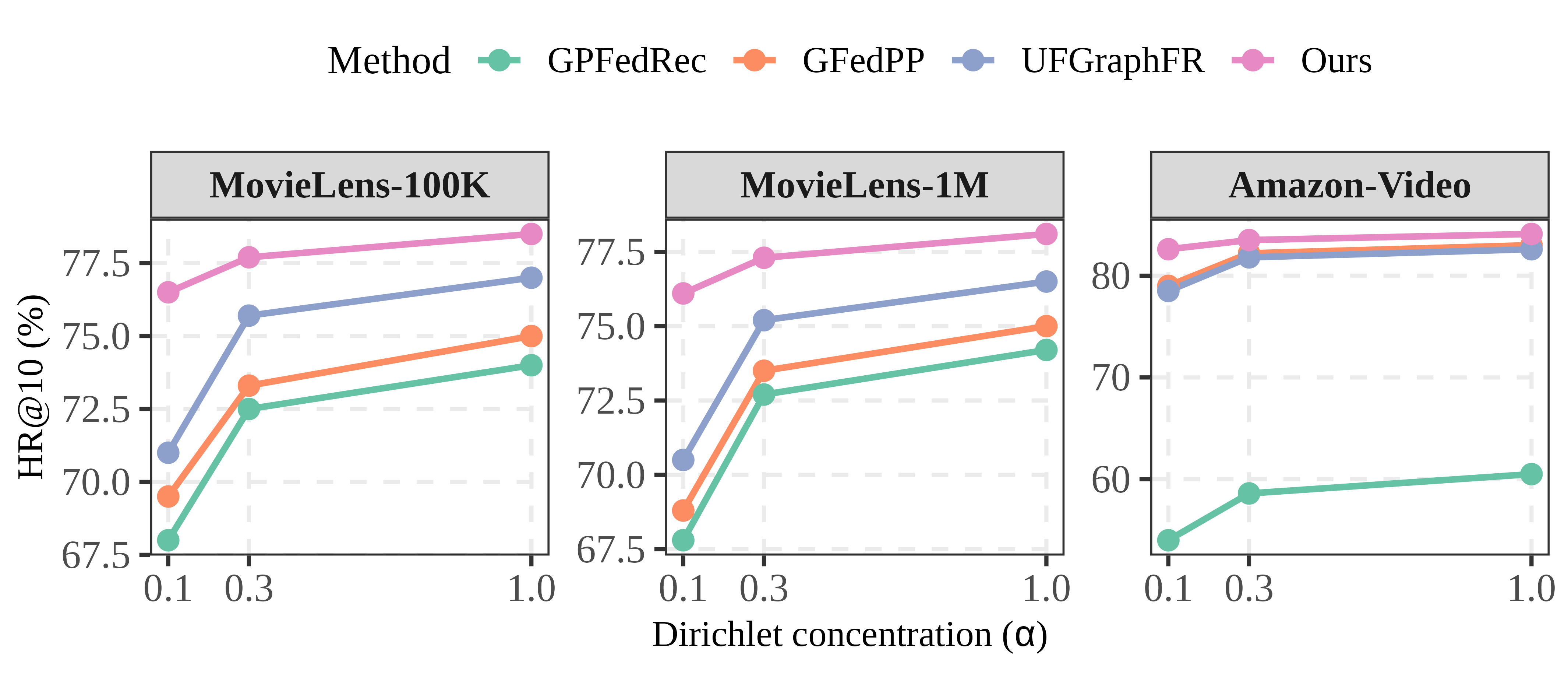}
    \caption{HR@10 (\%) under different levels of data heterogeneity 
    ($\alpha \in \{0.1, 0.3, 1.0\}$). Smaller $\alpha$ indicates more severe non-IID distributions.}
    \label{fig:noniid_results}
\end{figure*}

Figure~\ref{fig:noniid_results} presents HR@10 under varying degrees of client heterogeneity. 
As $\alpha$ decreases from 1.0 to 0.1, representative baselines such as UFGraphFR drop notably (e.g., 77.0\% to 71.0\% on MovieLens-100K), indicating sensitivity to non-IID distributions. 
In contrast, \toolname shows much smaller degradation: 78.5\% to 76.5\% on MovieLens-100K, and 78.1\% to 76.1\% on MovieLens-1M. It consistently maintains the highest performance across all $\alpha$ values. 
These results demonstrate that the semantic-first merging strategy effectively stabilizes cross-client aggregation under heterogeneous settings.

\subsection{RQ3: Ablation Study}

\begin{figure*}[!t]
    \centering
    \includegraphics[width=\textwidth]{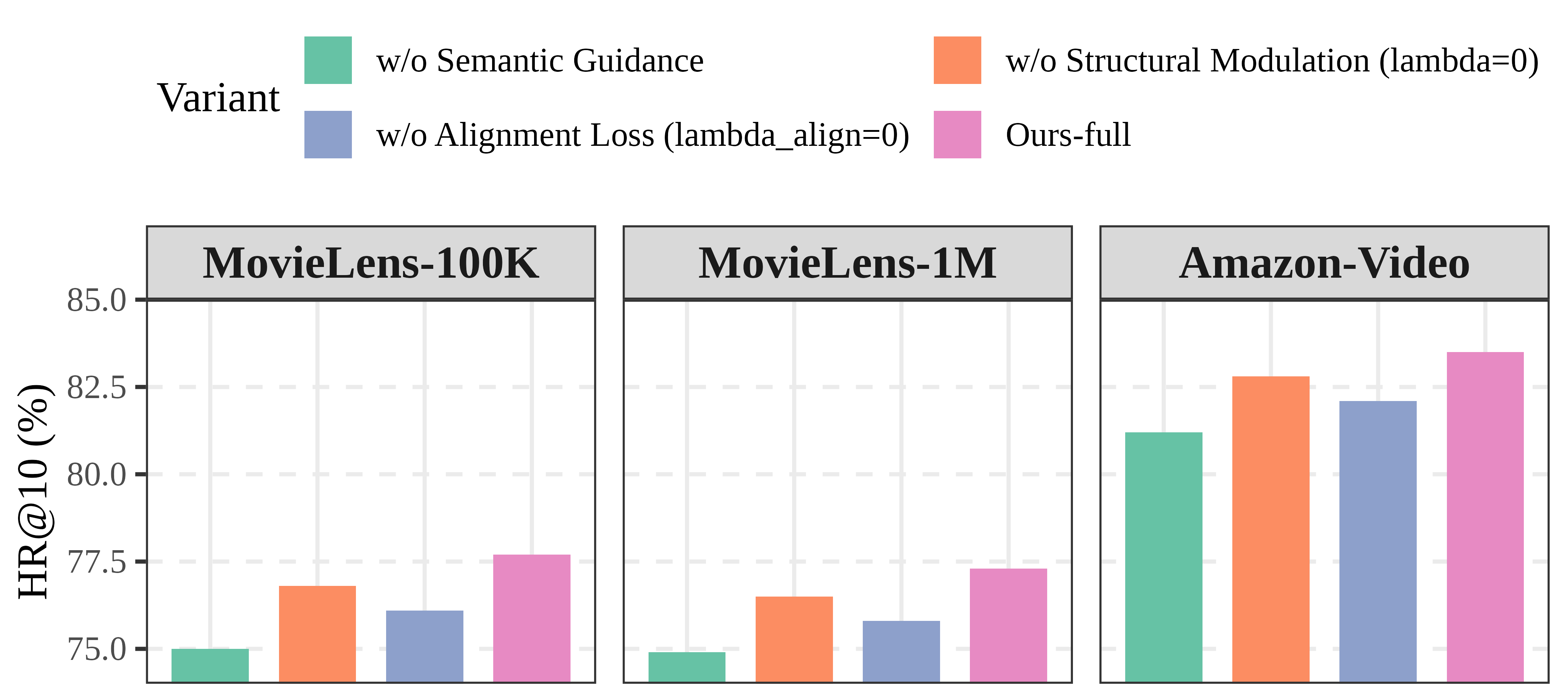}
      \caption{Ablation study on HR@10 (\%) under the non-IID setting ($\alpha=0.3$).}
    \label{fig:ablation}
\end{figure*}

Figure~\ref{fig:ablation} reports the contribution of each component under the non-IID setting ($\alpha=0.3$). 
Removing semantic guidance leads to the largest performance drop across all datasets; for example, HR@10 decreases from 77.7\% to 75.0\% on MovieLens-100K and from 83.5\% to 81.2\% on Amazon-Video. 
Disabling the alignment loss also degrades performance (e.g., 77.7\%→76.1\% on MovieLens-100K), while removing structural modulation (\(\lambda_{\text{merge}}=0\)) results in a smaller but consistent reduction. 
These results indicate that semantic-first merging is the most critical component, and that both structural modulation and alignment-aware optimization contribute to the overall effectiveness of the framework.

\subsection{RQ4: Hyperparameter Sensitivity}

\begin{figure*}[!t]
    \centering
    \includegraphics[width=\textwidth]{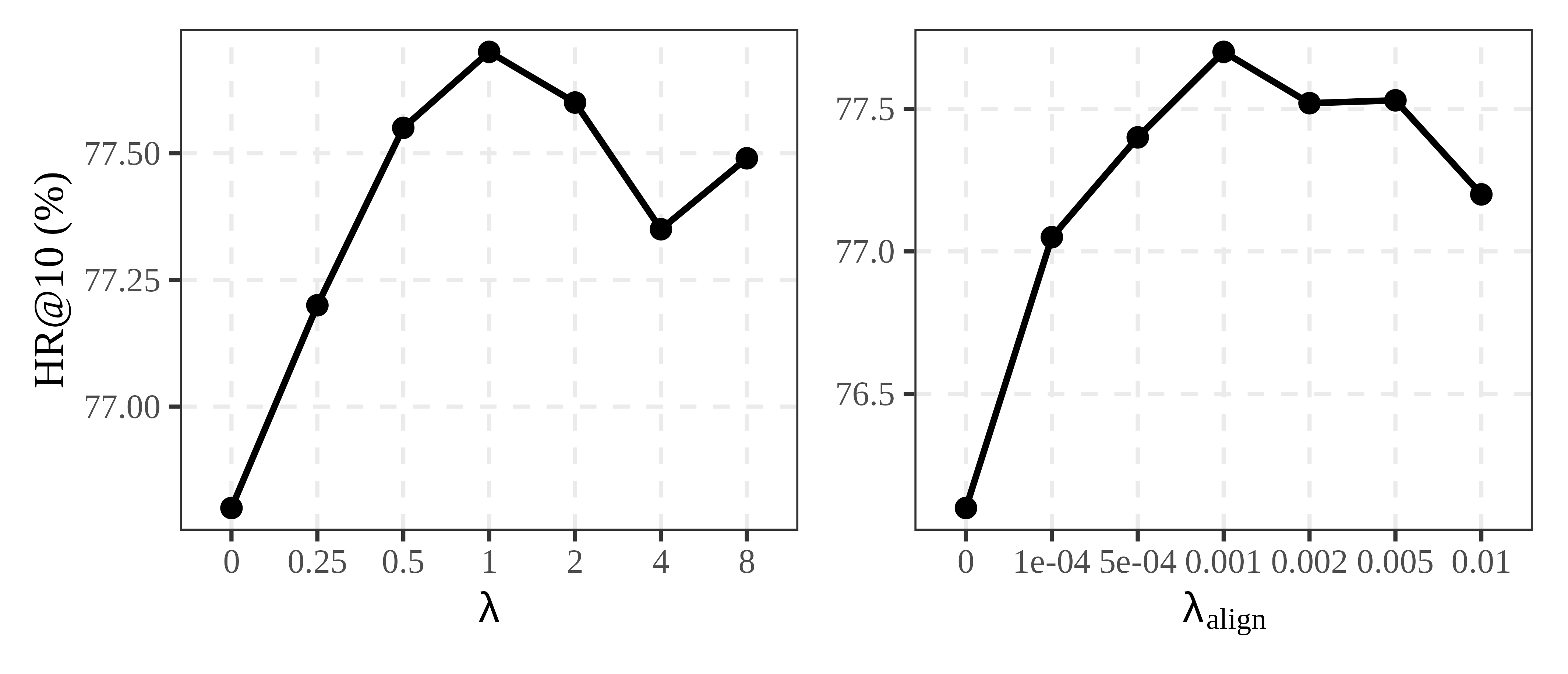}
    \caption{Sensitivity analysis of the structural modulation weight $\lambda$ and the alignment weight $\lambda_{\text{align}}$ on MovieLens-100K under the non-IID setting ($\alpha=0.3$), measured by HR@10 (\%).}
    \label{fig:sensitivity}
\end{figure*}

Figure~\ref{fig:sensitivity} shows that \toolname is stable across a broad range of hyperparameter values. 
For the structural modulation weight, performance improves from 76.8\% at $\lambda=0$ to a peak of 77.7\% at $\lambda=1$, and remains competitive even for larger values (e.g., 77.49\% at $\lambda=8$). 
Similarly, the alignment weight exhibits a clear sweet spot: HR@10 rises from 76.1\% at $\lambda_{\text{align}}=0$ to 77.7\% at $\lambda_{\text{align}}=10^{-3}$, and varies only mildly thereafter (e.g., 77.53\% at $\lambda_{\text{align}}=5\times10^{-3}$). 
Overall, the method is not overly sensitive to tuning, indicating robust behavior with respect to key hyperparameters.

\section{Conclusion}
\label{sec:conclusion}

\begin{sloppypar}
This paper presents a semantic--structural federated graph recommendation framework to mitigate representation misalignment under non-IID client distributions. Instead of relying solely on structural parameter aggregation, the proposed approach introduces community-level abstraction and semantic-first merging to regulate cross-client collaboration. By constructing structural prototypes from local interaction graphs and guiding their alignment using LLM-derived semantic representations, the framework enables effective knowledge transfer while keeping raw interaction data local. Experimental results demonstrate consistent improvements over strong federated and graph-based baselines under heterogeneous settings, and confirm the stability of the method across different hyperparameter configurations.

For future work, we plan to conduct a comprehensive system-level evaluation of the framework, including communication cost per round, runtime performance, and CPU/GPU resource usage under different deployment settings. We also aim to explore large-scale distributed deployment scenarios to better understand scalability and system behavior under realistic network and client heterogeneity. In addition, dynamic semantic updating strategies and more efficient community representation mechanisms will be investigated to further improve adaptability and system efficiency.
\end{sloppypar}



\newpage

\begin{thebibliography}{10}
\providecommand{\url}[1]{\texttt{#1}}
\providecommand{\urlprefix}{URL }
\providecommand{\doi}[1]{https://doi.org/#1}

\bibitem{ammad2019federated}
Ammad-Ud-Din, M., Ivannikova, E., Khan, S.A., Oyomno, W., Fu, Q., Tan, K.E.,
  Flanagan, A.: Federated collaborative filtering for privacy-preserving
  personalized recommendation system. arXiv  (2019)

\bibitem{campello2013density}
Campello, R.J., Moulavi, D., Sander, J.: Density-based clustering based on
  hierarchical density estimates. In: PAKDD. pp. 160--172 (2013)

\bibitem{chai2020secure}
Chai, D., Wang, L., Chen, K., Yang, Q.: Secure federated matrix factorization.
  IEEE Intelligent Systems  \textbf{36}(5),  11--20 (2020)

\bibitem{chen2025defedgcn}
Chen, Q., Wang, Z., Yan, M., Yan, H., Lin, X., Zhou, J.: Defedgcn:
  privacy-preserving decentralized federated gcn for recommender system. TSC
  \textbf{18}(2),  729--742 (2025)

\bibitem{harper2015movielens}
Harper, F.M., Konstan, J.A.: The movielens datasets: History and context. ACM
  TiiS  \textbf{5}(4),  1--19 (2015)

\bibitem{he2020lightgcn}
He, X., Deng, K., Wang, X., Li, Y., Zhang, Y., Wang, M.: Lightgcn: Simplifying
  and powering graph convolution network for recommendation. In: SIGIR. pp.
  639--648 (2020)

\bibitem{he2017neural}
He, X., Liao, L., Zhang, H., Nie, L., Hu, X., Chua, T.S.: Neural collaborative
  filtering. In: WWW. pp. 173--182 (2017)

\bibitem{koren2009matrix}
Koren, Y., Bell, R., Volinsky, C.: Matrix factorization techniques for
  recommender systems. Computer  \textbf{42}(8),  30--37 (2009)

\bibitem{kumar2025depth}
Kumar, C., Alam, M.: An in-depth analysis of recommender systems for
  integration of knowledge graphs utilising federated xai model. International
  Journal of Information Technology  \textbf{17}(5),  3147--3155 (2025)

\bibitem{liu2022federated}
Liu, Z., Yang, L., Fan, Z., Peng, H., Yu, P.S.: Federated social recommendation
  with graph neural network. TIST  \textbf{13}(4),  1--24 (2022)

\bibitem{long2024physical}
Long, J., Chen, T., Ye, G., Zheng, K., Nguyen, Q.V.H., Yin, H.: Physical
  trajectory inference attack and defense in decentralized poi recommendation.
  In: WWW. pp. 3379--3387 (2024)

\bibitem{long2024diffusion}
Long, J., Ye, G., Chen, T., Wang, Y., Wang, M., Yin, H.: Diffusion-based
  cloud-edge-device collaborative learning for next poi recommendations. In:
  KDD. pp. 2026--2036 (2024)

\bibitem{llmrec}
Lyu, H., Jiang, S., Zeng, H., Xia, Y., Wang, Q., Zhang, S., Chen, R., Leung,
  C., Tang, J., Luo, J.: {LLM}-rec: Personalized recommendation via prompting
  large language models. In: NAACL. pp. 583--612 (2024)

\bibitem{ma2024fedkgrec}
Ma, X., Zhang, H., Zeng, J., Duan, Y., Wen, X.: Fedkgrec: privacy-preserving
  federated knowledge graph aware recommender system: X. ma et al. Applied
  Intelligence  \textbf{54}(19),  9028--9044 (2024)

\bibitem{mai2023vertical}
Mai, P., Pang, Y.: Vertical federated graph neural network for recommender
  system. In: ICML. pp. 23516--23535 (2023)

\bibitem{na2025graph}
Na, C., Yang, K., Fang, D., Li, Y., Gao, J., Zhu, C., Zhang, J., Sun, X.,
  Chang, Y.: Graph federated learning for personalized privacy recommendation.
  arXiv  (2025)

\bibitem{ni2019justifying}
Ni, J., Li, J., McAuley, J.: Justifying recommendations using distantly-labeled
  reviews and fine-grained aspects. In: EMNLP-IJCNLP. pp. 188--197 (2019)

\bibitem{perifanis2022federated}
Perifanis, V., Efraimidis, P.S.: Federated neural collaborative filtering. KBS
  \textbf{242},  108441 (2022)

\bibitem{qi2025pfgrs}
Qi, Q., Hu, C., Li, T., Tang, P., Guo, S.: Pfgrs: a privacy-preserving
  subgraph-level federated graph learning for recommender system. ESWA
  \textbf{282},  127615 (2025)

\bibitem{qu2024towards}
Qu, L., Yuan, W., Zheng, R., Cui, L., Shi, Y., Yin, H.: Towards personalized
  privacy: User-governed data contribution for federated recommendation. In:
  WWW. pp. 3910--3918 (2024)

\bibitem{singhal2021federated}
Singhal, K., Sidahmed, H., Garrett, Z., Wu, S., Rush, J., Prakash, S.:
  Federated reconstruction: Partially local federated learning. NeurIPS
  \textbf{34},  11220--11232 (2021)

\bibitem{soyarar2025explaining}
Soyarar, E., Aydogan, R., Buzcu, B., Calvaresi, D.: Explaining federated
  learning-based movie recommendations. In: MetroXRAINE. pp. 729--734. IEEE
  (2025)

\bibitem{tan2025nah}
Tan, G.: Nah-gnn: A graph-based framework for multi-behavior and high-hop
  interaction recommendation. PloS one  \textbf{20}(4),  e0321419 (2025)

\bibitem{tang2025}
Tang, Z., Huan, Z., Li, Z., Zhang, X., Hu, J., Fu, C., Zhou, J., Zou, L., Li,
  C.: One model for all: Large language models are domain-agnostic
  recommendation systems. TOIS  \textbf{43}(5),  1--27 (2025)

\bibitem{wang2025ufgraphfr}
Wang, X., Hao, Q., Cheng, X., Xiao, Y.: Ufgraphfr: Graph federation
  recommendation system based on user text description features. arXiv  (2025)

\bibitem{wu2022federated}
Wu, C., Wu, F., Lyu, L., Qi, T., Huang, Y., Xie, X.: A federated graph neural
  network framework for privacy-preserving personalization. Nature
  Communications  \textbf{13}(1), ~3091 (2022)

\bibitem{wu2024llm4rec_survey}
Wu, L., Zheng, Z., Qiu, Z., Wang, H., Gu, H., Shen, T., Qin, C., Zhu, C., Zhu,
  H., Liu, Q., Xiong, H., Chen, E.: A survey on large language models for
  recommendation. WWW  \textbf{27}(5), ~60 (2024)

\bibitem{xu2025enhancing}
Xu, Z., Li, B., Cao, W.: Enhancing federated learning-based social
  recommendations with graph attention networks. Neurocomputing  \textbf{617},
  129045 (2025)

\bibitem{yang2024federated}
Yang, Z., Peng, Z., Wang, Z., Qi, J., Chen, C., Pan, W., Wen, C., Wang, C.,
  Fan, X.: Federated graph learning for cross-domain recommendation. NeurIPS
  \textbf{37},  64865--64888 (2024)

\bibitem{yao2024}
Yao, Y., Zhang, J., Wu, J., Huang, C., Xia, Y., Yu, T., Zhang, R., Kim, S.,
  Rossi, R., Li, A., Yao, L., McAuley, J., Chen, Y., Joe-Wong, C.: Federated
  large language models: Current progress and future directions (2024)

\bibitem{ye2024}
Ye, R., Ge, R., Zhu, X., Chai, J., Du, Y., Liu, Y., Wang, Y., Chen, S.:
  {FedLLM-Bench}: Realistic benchmarks for federated learning of large language
  models (2024)

\bibitem{yin2025device}
Yin, H., Qu, L., Chen, T., Yuan, W., Zheng, R., Long, J., Xia, X., Shi, Y.,
  Zhang, C.: On-device recommender systems: A comprehensive survey. Data
  Science and Engineering pp. 1--30 (2025)

\bibitem{yuan2024hetefedrec}
Yuan, W., Qu, L., Cui, L., Tong, Y., Zhou, X., Yin, H.: Hetefedrec: Federated
  recommender systems with model heterogeneity. In: ICDE. pp. 1324--1337 (2024)

\bibitem{yuan2024hide}
Yuan, W., Yang, C., Qu, L., Nguyen, Q.V.H., Li, J., Yin, H.: Hide your model: A
  parameter transmission-free federated recommender system. In: ICDE. pp.
  611--624 (2024)

\bibitem{yuan2025fellas}
Yuan, W., Yang, C., Ye, G., Chen, T., Nguyen, Q.V.H., Yin, H.: Fellas:
  Enhancing federated sequential recommendation with llm as external services.
  TOIS  \textbf{43}(6),  1--24 (2025)

\bibitem{yurochkin2019bayesian}
Yurochkin, M., Agarwal, M., Ghosh, S., Greenewald, K., Hoang, N., Khazaeni, Y.:
  Bayesian nonparametric federated learning of neural networks. In: ICML. pp.
  7252--7261 (2019)

\bibitem{zeng2024federated}
Zeng, H., Yue, Z., Jiang, Q., Wang, D.: Federated recommendation via hybrid
  retrieval augmented generation. In: BigData. pp. 8078--8087. IEEE (2024)

\bibitem{zhang2023dual}
Zhang, C., Long, G., Zhou, T., Yan, P., Zhang, Z., Zhang, C., Yang, B.: Dual
  personalization on federated recommendation. arXiv  (2023)

\bibitem{zhang2024gpfedrec}
Zhang, C., Long, G., Zhou, T., Zhang, Z., Yan, P., Yang, B.: Gpfedrec:
  Graph-guided personalization for federated recommendation. In: KDD. pp.
  4131--4142 (2024)

\bibitem{zhao2025federated}
Zhao, J., Wang, W., Xu, C., Ng, S.K., Chua, T.S.: A federated framework for
  llm-based recommendation. In: NAACL. pp. 2852--2865 (2025)

\end{thebibliography}

\end{document}